\documentclass[aps,showpacs,preprint]{revtex4}
\usepackage{graphicx}
\def\>{\rangle}

\def\ket{\rangle}
\def\<{\langle}
\def\>{\rangle}
\begin{document}
\title{Nuclear Magnetic
Resonance Implementation of a Quantum Clock Synchronization
Algorithm \footnote{Corresponding authors:\\ Jingfu Zhang,
zhang-jf@mail.tsinghua.edu.cn and Gui Lu Long,
gllong@mail.tsinghua.edu.cn}}
\author{ Jingfu Zhang$^{1,2}$, Gui Lu Long$^{1,2}
$, Zhiwei Deng,$^{3}$, Wenzhang Liu$^{1,2}$
     and Zhiheng Lu,$^{4}$}
\address{ $^{1}$Key Laboratory For Quantum Information and Measurements,  and
Department of Physics, Tsinghua University, Beijing,
100084, P R China\\
 $^{2}$Center For Quantum Information, Tsinghua University, Beijing 100084, P R China\\
$^{3}$Testing and Analytical Center, Beijing Normal University,
 Beijing, 100875, P R China\\
$^{4}$Department of Physics, \small{}Beijing Normal University,
Beijing, 100875, P R China }
\date{\today}
\begin{abstract}
 The quantum clock synchronization (QCS) algorithm proposed by
Chuang (Phys. Rev. Lett, 85, 2006(2000)) has been implemented in a
three qubit nuclear magnetic resonance quantum system. The time
difference between two separated clocks can be determined by
measuring the output states. The experimental realization of the
QCS algorithm also demonstrates an application of the quantum
phase estimation.
\end{abstract}
 \maketitle

\section{Introduction}
\label{s1}

   The combination of quantum mechanics and computer science gives birth to quantum
   computer where    quantum properties enable quantum computer to efficiently solve
   difficult problems in classical computer,
   for instance to factorize a large number using the Shor algorithm\cite{shor}.
    Marriage of quantum mechanics with other traditional science and
   technology has also produced fruitful results.
 Clock synchronization is one such example. The task of clock synchronization is to
   determine the time difference
$\Delta$ between two spatially separated clocks. It is  important
both in practical application and in scientific research.
Recently, some authors have applied quantum mechanical means to
this classical problem and showed that quantum clock
synchronization gains significant improvement compared to its
classical counterpart\cite{Chuang,Jozsa,Giovannetti}. One of the
quantum clock synchronization(QCS) algorithms is the one proposed
by Chuang \cite{Chuang}, which uses the quantum phase estimation
method\cite{phas1,phas2}. Chuang's QCS algorithm  obtains $n$
digits of accuracy in the time difference $\Delta$ while
exchanging only the order of $n$ number of  qubits. This quantum
algorithm gains an exponential improvement over classical
algorithms which requires $O(2^{n})$ message exchanges.

With the highest speed, photon is the natural choice for the
practical implementation of the QCS algorithm.   An alternative
quantum system to implement the QCS algorithm is to use quantum
spins in a magnetic field as suggested in Ref. \cite{Chuang}.
 Nuclear magnetic resonance (NMR) has been widely used in various fields,
 and it has also become an
important arena to demonstrate quantum algorithms. Many
algorithms, such as the Grover algorithm, the quantum Fourier
transform, and Shor's quantum factoring algorithm, have been
demonstrated in NMR quantum
systems\cite{Chuangprl,Vandersypen011,Weinstein,Long02,Peng}. Some
quantum communication protocols, such as the quantum
teleportation, quantum dense coding have also been demonstrated in
NMR quantum systems\cite{tele,Fang,dense}. It is thus a good
system to demonstrate the QCS algorithm.

  In this paper, we report the result of an implementation of
the QCS algorithm in a three qubit NMR quantum computer. The basis
of the algorithm is the quantum Fourier transform (QFT) which has
been applied in experiments previously
\cite{Weinstein,Lieven,Weinstein02}. The QCS algorithm requires an
pure state as the initial state. To use an NMR ensemble for
quantum computation, preparation of the effective-pure state is
one necessary step\cite{cory97,gersh97}. Temporal averaging and
spatial averaging are two main practical methods to prepare
effective-pure states \cite{s12,Cory,Knill1}. In our experiments,
we chose the spatial averaging method to prepare the
effective-pure state. Using this method, the algorithm is
implemented through only one experiment. Compared with the
temporal averaging, the spatial averaging can shorten experiment
time, and it has been applied in various experiments
\cite{Somarooprl991,Kill00,Cory98,Boulant,Teklemariam,Boulant03,Viola}.

This paper is organized as follows. After this brief introduction,
we give the outline of the QCS algorithm, and in particular we
give the explict quantum circuit of the QCS algorithm in a three
qubit system in section \ref{s2}. In section \ref{s3}, we give the
details of the experimental demonstration of the QCS algorithm,
including the pulse sequence and the results of the experiment. In
section \ref{s4}, we give a brief summary.

\section{The QCS Algorithm and the Simplified Circuit for a 3-qubit NMR Quantum System}
\label{s2}

 A brief summary of the QCS algorithm is given below,
and the details can be found in Ref.\cite{Chuang}.  In the QCS
algorithm, Alice possesses $m+1$ qubits. The first $m$ qubits are
the working qubits and are retained at Alice's site, and the extra
qubit is an ancilla qubit for implementing the $T$-unitary
operation. The procedure of the QCS is as follows. Alice performs
the Hadamard operation on each of the first $m$ qubits. This
prepares the state of the $m+1$ qubit system to
$$|\phi_1\ket={1\over \sqrt{2^m}}\sum_{k=0}^{2^m-1}|k\ket|0\ket.$$
 Then the $T$ operation, which inscribe the time difference between
 the two clocks into the many-body quantum state,
 is implemented. At this stage the state of $m+1$ qubit system is
$$|\phi_2\ket={1\over\sqrt{2^m}}\sum_{k=0}^{2^m-1} e^{2\pi i k
\omega \Delta}|k\ket |0\ket.
$$ Then Alice applies an inverse quantum Fourier transform
$F^{-1}$ on the first $m$ qubits, transforming their state to
\begin{eqnarray}
|\phi_3\ket={1\over
2^m}\sum_{k=0}^{2^m-1}\sum_{j=0}^{2^m-1}e^{2\pi i
k(\omega\Delta-j/2^m)}|j\ket=\sum_{j=0}^{2^m-1}c_j|j\ket.
\end{eqnarray}
The state is peaked at $j=2^m\omega\Delta$. If $2^m\omega \Delta$
is an integer, the equality is exact. Measuring the qubit values
of the first $m$ qubits gives us the value of $j$ which in turn
determines the time difference $\Delta$. In practice, the equality
can not hold exactly, and it gives $\omega\Delta$ about $m$ bits
of accuracy with a high probability.

 The central ingredients of the QCS algorithm are
  the $T$ operation and the quantum Fourier transform. The $T$ operation
 can be implemented  by performing the following operation on each of the $m$ qubits:
1) Alice makes CNOT operation on the $l$-th and the ancilla qubit
which is the target state ; 2) The ticking qubit handshake
protocol ${\rm TQH}(\pi 2^{l-1}\omega,|\psi\ket)$ is performed,
transforming the state of the $l$-th and ancilla qubit system into
\begin{eqnarray}
{1\over \sqrt{2}}\left(e^{-2^l\pi i\omega
\Delta/2}|00\ket+e^{2^l\pi i \omega \Delta/2}|11\ket\right).
\end{eqnarray}
3) Alice performs another CNOT gate to the $l$-th and the ancila
qubit, so that the state of the $l$-th and the ancilla qubit
becomes
\begin{eqnarray}
e^{-2^l\pi i\omega \Delta /2}\sum_{k_l=0}^1{1\over \sqrt{2}}
|k_l\ket e^{2^l\pi i k_l\omega \Delta}|0\ket.
\end{eqnarray}
4) After all the $m$ qubit have gone through the previous
operations, the overall operation is to transform the state
$|k^{\prime}\ket |0\ket=|k_0 k_1 \cdots k_{m-1}\ket |0\ket$ into
$|k^{\prime}\ket e^{2\pi i \omega\Delta (\sum_l2^lk_l)}|0\ket$,
because
\begin{eqnarray}\label{T}
T|k^{\prime}\ket|0\ket =[T_0|k_0\ket T_1|k_1\ket \cdots
T_{m-1}|k_{m-1}\ket]|0\ket=|k^{\prime}\ket e^{2\pi i \omega\Delta
(\sum_l2^lk_l)}|0\ket.
\end{eqnarray}
Using the SWAP operation,  the state $|k^{\prime}\ket$ is
transformed into $|k\ket=|k_{m-1}\cdots  k_1 k_0\ket$, and Eq.
(\ref{T}) becomes
\begin{eqnarray}
|k\ket e^{2\pi i k\omega \Delta }|0\ket,
\end{eqnarray}
where $\sum_l2^lk_l=k$.

\begin{figure}
\includegraphics[width=6in]{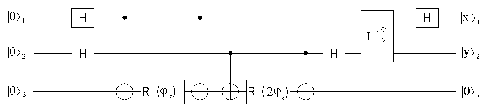}
\caption{The quantum network to implement the QCS algorithm. The
three lines denote the three qubits. $H$ denotes the Hadmamard
transform, $I_{|11\rangle}^{-\pi/2}$ denotes the controlled phase
shift operation, and $R_{z}(\varphi_{k})=e^{i\varphi_{k}I_{z}}$,
where $\varphi_{k}=-k\pi/2$ $(k=0,1,2,3)$.
 The network outlined by the dashed rectangle implements the inverse of quantum Fourier
transform without the SWAP operation. Time goes from left to
right. $|x\rangle_{1}|y\rangle_{2}$ is the output state, which can
be obtained through measuring  qubits 1 and 2.} \label{network}
\end{figure}

The quantum network shown in Fig. \ref{network} implements the QCS
algorithm in a three qubit system. The three lines denote the
three qubits respectively. $|0\rangle$ denotes the spin up state.
$H$ denotes the Hadamard transform. The effect of the TQH is to
introduce a phase to the two different quantum states of the
ancilla qubit, namely for the $|0\ket$ state with phase
$e^{-2^l\pi i \omega \Delta/2}$ and for state $|1\ket$with phase
$e^{2^l\pi i \omega \Delta/2}$, hence the state ${1\over
\sqrt{2}}(|00\ket+|11\ket)$ of the $l$-th qubit and the ancilla
qubit system changes to
\begin{eqnarray}
{1\over \sqrt{2}}\left(e^{-2^l\pi i \omega
\Delta/2}|00\ket+e^{2^l\pi i \omega \Delta/2}|11\ket\right).
\end{eqnarray}
In NMR, this is equivalent to a rotation about the $z$-axis. For
the first qubit, the rotation is
 $R_{z}(\varphi_{k})=e^{i\varphi_{k}I_{z}}$, and for the second
 qubit the rotation is
$R_{z}(2\varphi_{k})=R_{z}(\varphi_{k})R_{z}(\varphi_{k})$. We
have written $-\omega\Delta\pi$ as $\varphi$. Because
$2^2\omega\Delta$ has to be integer, ranging from 1 to $2^m-1=3$,
so $\varphi$ can take the following values $\varphi_{k}=-k\pi/2$,
where $k=0,1,2,3$. When $m$ is large, $k$ takes the value from 0
to $2^m-1$ and the measurement of this $k$ value gives the value
of $\omega\Delta$, and hence the time difference $\Delta$.

For the quantum Fourier transform, some simplification is
possible. We use $I_{|11\rangle}^{-\pi/2}$ to denote the
controlled phase shift operation applied to the subsystem
constructed by qubit 1 and 2. $I_{|11\rangle}^{-\pi/2}$ is
explicitly
\begin{equation}\label{shift}
 I_{|11\rangle}^{-\pi/2}=\left( \begin{array}{cccc}
   1&0 &0 & 0 \\
   0 & 1 & 0 & 0 \\
   0 & 0 & 1 & 0 \\
   0 & 0 & 0 & -i \\
 \end{array}\right),
\end{equation}
with the basis orders in $|00\rangle$, $|01\rangle$, $|10\rangle$,
$|11\rangle$. The network outlined by the dashed rectangle in
Fig.\ref{network} implements the inverse of the quantum fourier
transform, where the SWAP operation has been counteracted by
another one in the network. The inverse of the QFT $F^{-1}$ can be
written as $F^{\prime-1}$SWAP\cite{Weinstein}. The effect of this
network is to make the following transformation
\begin{equation}\label{fprime}
 F^{\prime-1}=\frac{1}{2}\left(\begin{array}{cccc}
   1&1 &1 & 1 \\
   1 & -1 & -i & i \\
   1 & 1 & -1 & -1 \\
   1 & -1 & i & -i \\
 \end{array}\right).
\end{equation}
The bit values of the first two qubits are the desired output. The
network shown in Fig. \ref{network} transforms $|000\rangle$ to
$|000\rangle$, $|010\rangle$, $|100\rangle$, and $|110\rangle$,
corresponding to $\varphi_{k}=0$, $-\pi/2$, $-\pi$, and $-3\pi/2$,
respectively, and $\omega\Delta$ to be 0, 1/4, 2/4 and 3/4
respectively.  By measuring qubit 1 and 2, one obtains the
concrete value of $\varphi$ and hence determines the time
difference between the two clocks.

\section{Implementation in a 3-qubit NMR quantum system}
\label{s3}

\begin{figure}
\includegraphics[width=4.5in]{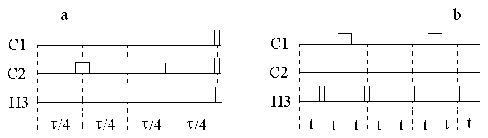}
\caption{Pulse sequence to realize $[\tau]_{13}$ (a) and the
chemical shift evolution $R^{2}_{z}(2\pi \delta 8t)$ (b), where
$\delta$ is the offset between the chemical shift of C2 and the
transmitter frequency. Rectangles denote $\pi$ pulses. The narrow
ones are so short that the widths can be ignored. The wide ones,
however, are long pulses of which widths can not be ignored.}
\label{j13}
\end{figure}

   The experiment uses a sample of Carbon-13 labelled
trichloroethylene (TCE) dissolved in d-chloroform. Data are taken
at room temperature with a Bruker DRX 500 MHz spectrometer.
$^{1}$H is denoted as qubit 3, the $^{13}$C directly connecting to
$^{1}$H is denoted as qubit 2, and the other $^{13}$C is denoted
as qubit 1. The three qubits are denoted as C1, C2 and H3. By
setting $\hbar=1$, the Hamitonian of the three-qubit system is
\cite{s10}
\begin{equation}\label{hamidun}
  H=-2\pi\nu_{1}I_{z}^{1}-2\pi\nu_{2}I_{z}^{2}-2\pi\nu_{3}I_{z}^{3}
  +2\pi J_{12}I_{z}^{1}I_{z}^{2}+2\pi J_{23 }I_{z}^{2}I_{z}^{3}
  +2\pi J_{13} I_{z}^{1}I_{z}^{3}.
\end{equation}
$I_{z}^{j}(j=1,2,3)$ are the matrices for $z$-component of the
angular momentum of the spins. $\nu_{1}$, $\nu_{2}$, $\nu_{3}$ are
the resonance frequencies of C1, C2 and H3, and
$\nu_{1}=\nu_{2}+904.4$Hz. The coupling constants are measured to
be $J_{12}=103.1$ Hz, $J_{23}=203.8$ Hz, and $J_{13}=9.16$ Hz. The
coupled-spin evolution between two spins is denoted as
\begin{equation}\label{2}
  [\tau]_{jl}=e^{-i2\pi J_{jl} \tau I_{z}^{j} I_{z}^{l}},
\end{equation}
where $l=1,2,3$, and $j\neq l$. $[\tau]_{jl}$ can be realized by
averaging the coupling constants other than $J_{jl}$ to
zero\cite{s15}. For example, $[\tau]_{13}$ is realized by the
pulse sequence shown in Fig. \ref{j13} (a). The chemical shift
evolution of C2 is realized by the pulse sequence shown in Fig.
\ref{j13} (b).  $R^{2}_{z}(\pi)$ can be realized by choosing the
proper evolution time, and the transmitter frequency\cite{Linden}.
The $\pi$ pulses for C2 are chosen as RE-BURP pulses to excite the
multiplet of C2 uniformly\cite{Geen}.

The initial effective-pure state $|000\rangle$ is prepared by
spatial averaging\cite{Cory}.
   The following radio-frequency (rf) pulse and gradient pulse
sequence
\begin{eqnarray}\label{puls}
&&[\pi/4]_{x}^{1,2}-[1/2J_{12}]_{12}-[-5\pi/6]_{y}^{1,2}-[\alpha]_{x}^{3}-[{\rm
grad}]_{z}
-[\pi/4]_{y}^{3}-[9/2J_{23}]_{23}-\nonumber\\
&&[1/2J_{13}]_{13}-[\pi/4]_{y}^{3}-[{\rm grad}]_{z}
-[\pi/4]_{y}^{3}-[9/4J_{23}]_{23}-[1/4J_{13}]_{13}-[\pi/4]_{x}^{3}-[{\rm
grad}]_{z},
\end{eqnarray}
transforms the system from the equilibrium state
\begin{equation}\label{equ}
  \rho_{eq}=\gamma_{C}(I_{z}^{1}+ I_{z}^{2})+\gamma_{H}I_{z}^{3},
\end{equation}
to
\begin{equation}\label{2}
  \rho_{0}=I_{z}^{1}/2+I_{z}^{2}/2+I_{z}^{3}/2+I_{z}^{1}I_{z}^{2}+I_{z}^{2}I_{z}^{3}+
  I_{z}^{1}I_{z}^{3}+2I_{z}^{1}I_{z}^{2}I_{z}^{3},
\end{equation}
where an overall phase factor has been
ignored\cite{Cory,Knill1,Somarooprl991,Zhang10}.
$[\pi/4]_{x}^{1,2}$ denotes the $\pi/4$ pulse exciting C1 and C2
simultaneously along x-axis. $[\pi/4]_{y}^{3}$ denotes the
spin-selective $\pi/4$ pulse for $^{1}$H along y-axis.
$\gamma_{C}$ and $\gamma_{H}$ denotes the gyromagnetic ratio of
$^{13}C$ and $^{1}$H.
$\alpha=\arccos(-\gamma_{C}\sqrt{6}/\gamma_{H})$. $\rho_{0}$ is
equivalent to $|000\rangle$. We find that the compound operation

\begin{equation}\label{c13}
 {\rm CNOT}_{13}R_{z}^{3}(\varphi_{k}){\rm CNOT}_{13}
=[\frac{-\varphi_{k}}{\pi J_{13}}]_{13},
\end{equation}

\begin{equation}\label{c23}
 {\rm CNOT}_{23}R_{z}^{3}(2\varphi_{k}){\rm CNOT}_{23}
=[\frac{-2\varphi_{k}}{\pi J_{23}}]_{23},
\end{equation}
realize the network much easier. The Hadamard transform
simultaneously applied to C1 and C2, is realized by pulse sequence
$[-\pi/2]_{y}^{1,2}-[\pi]_{x}^{1,2}$ where the number in the
superscript refers to the qubit number. The Hadamard transform for
C2 in the inverse of QFT, denoted by $H^{2}$, is realized  by
$[\pi/4]_{y}^{1,2,3}-R_{z}^{2}(\pi)-[-\pi/4]_{y}^{1,2,3}$, and
$H^{1}=H^{1,2}H^{2}$, noting $H^{-1}=H$. $I_{|11\rangle}^{-\pi/2}$
is realized by
$[1/4J_{12}]-[-\pi/2]^{1,2}_{y}-[\pi/4]^{1,2}_{x}-[\pi/2]^{1,2}_{y}
 $\cite{Weinstein,Longpla}.

\begin{figure}
\includegraphics[width=2.8in]{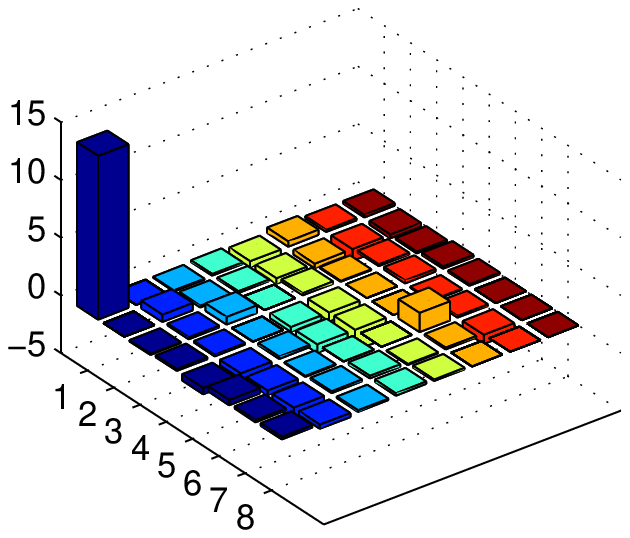}
\caption{The density matrix (in arbitrary units) reconstructed
using the state tomography technique, when the system is prepared
in effective- pure state $|000\rangle$.} \label{state}
\end{figure}

    The experimental results are represented as the density matrices
obtained by the state tomography technique
\cite{s12,nmr,Leskowitz}, where the spin-selective readout pulses
for C2, denoted as $[\pi/2]_{x}^{2}$ and $[\pi/2]_{y}^{2}$, are
realized by pulse sequences
$[\pi/2]_{y}^{1,2}-R_{z}^{2}(\pi/2)-[-\pi/2]_{y}^{1,2}$ and
$[-\pi/2]_{x}^{1,2}-R_{z}^{2}(\pi/2)-[\pi/2]_{y}^{1,2}$,
respectively, in order to freeze the motion of C1
\cite{Linden,Vandersypen}. $R_{z}^{2}(\pi/2)$ is realized by the
pulse sequence shown in Fig. \ref{j13}(b). Fig. \ref{state} shows
the experimentally measured density matrix when the system lies in
effective- pure state $|000\rangle$ prepared by the pulse sequence
(\ref{puls}). In the generated density matrix, the desired
element, which is the only nonzero element in theory, is measured
to be 14.2 (in arbitrary units). The amplitudes of the other
elements, which are zero in theory, are less than 1.8.

\begin{figure}
\includegraphics[width=6in]{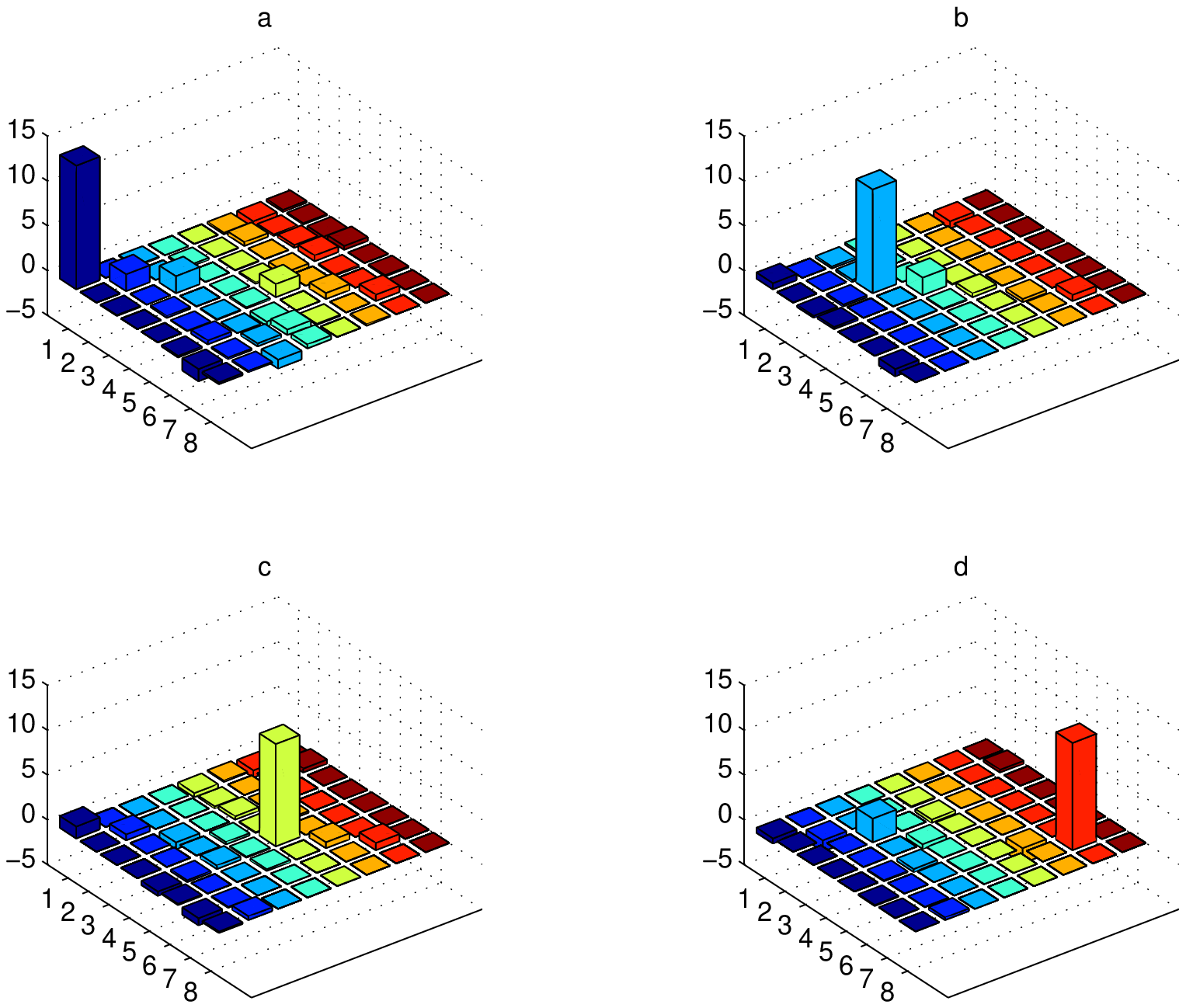}
\caption{The density matrices of the three- qubit system, after
the implementation of the QCS algorithm. Only the real components
are plotted. The imaginary portions, which are theoretically zero,
are found to contribute less than $10\%$ to the experimental
results. Figs. (a-d) correspond to the four effective- pure states
$|000\rangle$, $|010\rangle$, $|100\rangle$, and $|110\rangle$,
and the four effective- pure states correspond to the four
different time differences $\Delta=0$, $\Delta=1/4\omega$,
$\Delta=1/2\omega$, and $\Delta=3/4\omega$, respectively.  }
\label{states12}
\end{figure}

  The QCS algorithm starts with effective- pure
state$|000\rangle$. When $\varphi_{0}=0$, $\varphi_{1}=-\pi/2$,
$\varphi_{2}=-\pi$, and $\varphi_{3}=-3\pi/2$, the network shown
in Fig. \ref{network} transforms  $|000\rangle$ to $|000\rangle$,
$|010\rangle$, $|100\rangle$, and $|110\rangle$, respectively,
corresponding to the four different time differences $\Delta=0$,
$\Delta=1/4\omega$, $\Delta=1/2\omega$, and $\Delta=3/4\omega$.
Figs. \ref{states12} (a-d) show the experimentally measured
density matrices of the three- qubit system after the completion
of the  QCS algorithm, corresponding to
$\varphi_{0}$-$\varphi_{3}$, respectively. The fidelity of the
transformation is described by \cite{Weinstein}

\begin{equation}\label{acc}
    C=\frac{Tr(\rho_{theory}\rho_{exp})}{\sqrt{Tr(\rho^{2}_{theory})}\sqrt{Tr(\rho^{2}_{exp})}}
    \sqrt{\frac{Tr(\rho^{2}_{exp})}{Tr(\rho^{2}_{initial})}}.
\end{equation}
$\rho_{initial}$ is the initial density matrix shown in Fig.
\ref{state}. $\rho_{theory}=U \rho_{initial} U^{\dag}$, where $U$
denotes the theoretical transformation to implement the  QCS
algorithm. $\rho_{exp}$ denotes the experimentally measured
density matrix shown in Figs. \ref{states12}. The fidelities
corresponding to Figs. \ref{states12} (a-d) are $90.7\%$,
$77.4\%$, $75.2\%$, and $77.0\%$, respectively. The errors mainly
result from the imperfection of the pulses, the inhomogeneity in
the magnetic field and the decoherence time limit.

\section{Conclusion}
\label{s4}

    We have implemented the quantum clock synchronization
algorithm in a three- qubit NMR quantum computer. Using spatial
averaging, the algorithm is implemented through one experimental
run which is much shorter than the temporal averaging method. For
small qubit system such as three qubit used in this experiment,
the level of accuracy as good as the temporal averaging method.
 Compared with temporal averaging, process
of experiments is simplified greatly. The time difference can be
can be read out through the output state. In the experiments, we
have exploited the long range coupling between non-adjacent
nuclear spins, and it is found that it works well. In NMR samples,
long range coupling is a precious resource, and use of this long
range coupling in experimental realization should be made as much
as possible. On the other hand,  through optimizing network,  the
experimental operation difficulty can be reduced. In this
experiment, using a simplified network where some redundant
operations have been got rid of, we have shortened the total time
consumed by the whole algorithm, and hence also reduces the effect
of decoherence consequently. Although we have demonstrated the
algorithm  with only three qubit,  the techniques can be
generalized to more qubit system. It should be pointed out QCS is
one promising area of quantum information technology that may be
implemented in the future because it requires less number of qubit
than other quantum algorithms. For instance to achieve an accuracy
of $100$ps, the number of qubits required is only 34. Of course,
before QCS can be realistically used in practice, there should be
big advancement in the technology, for instance in the accuracy of
quantum gate operations, and the length of decoherence time.

\section*{Acknowledgement}
   This work is supported by the National Natural Science
Foundation of China under Grant No. 10374010, 60073009, 10325521,
the National Fundamental Research Program Grant No. 001CB309308,
the Hang-Tian Science Fund, the SRFDP program of Education
Ministry of China, and  China Postdoctoral Science Foundation.

\newpage


\end{document}